# Confined exciton polaron in $MoS_2$ on twisted-hBN

Garima Gupta[1,2], Mayank Chhaperwal[1], Pankaj Kumar[1], Pushkar Dasika[1], Kenji Watanabe[3], Takashi Taniguchi[4], Archana Raja[2], and Kausik Majumdar[1,*]

[1]*Department of Electrical Communication Engineering, Indian Institute of Science, Bangalore 560012, India*

[2]*Molecular Foundry, Lawrence Berkeley National Laboratory, Berkeley, CA 94720, USA*

[3] *Research Center for Functional Materials, National Institute for Materials Science, 1-1 Namiki, Tsukuba 305-044, Japan*

[4] *International Center for Materials Nanoarchitectonics, National Institute for Materials Science, 1-1 Namiki, Tsukuba 305-044, Japan*

[*]*Corresponding author, E-mail:* kausikm@iisc.ac.in

**Abstract:** The simple electrostatic picture of a trion is that of an excess charge inducing an exciton polarization and binding closer (farther) to the hole (electron) side of it. Trion formation can be forbidden when such spontaneous rearrangement of charges is not allowed by the application of external perturbation, such as electric field. Here we test this hypothesis experimentally using a non-monotonic electric field. We realize this scenario by imprinting the ferroelectric domains at the AA-stacked twisted-hBN (t-hBN) interface onto a monolayer of $MoS_2$ placed over it. The spatially varying in-plane electric field around the domain wall serves the dual purpose of (a) confining and polarizing the 2D exciton in the domain wall, and (b) depleting the free charge carriers from the domain wall. We observe a large quantized exciton splitting confirming strong exciton confinement in the domain wall. Forced by the confining potential, the electron side of the polarized exciton lies closer to the domain with accumulated free electrons, which should ideally prevent any trion formation. Contrary to the laid hypothesis, we observe signatures of quantized charged exciton emission, with an inter-level splitting that mimics the level-splitting of the quantized excitons. This paradox is explained using the many-body picture of exciton polaron, where a conduction band hole attractively binds the polarized exciton and the electron Fermi sea. The results provide a definitive way to unambiguously discern exciton polaron from trion.

An exciton surrounded by excess free charge carriers (say, electrons) forms a three-particle bound quasiparticle known as trions. To form a trion, the electric field of the excess electron polarizes the exciton, and the excess electron gets attached closer to the hole side of the polarized exciton[1]. It has been argued[2–4] that this three-particle trion picture holds when the density ($N$) of free electrons (with Fermi energy $E_F$) is low ($E_F \ll E_T$, where $E_T$ is the trion binding energy). Under large $N$, excitons interact with the Femi sea of electrons and form exciton polarons[1,5–7]. At high n-doping, where the Fermi energy lies inside the conduction band, an electron in the Fermi sea, on interaction with the exciton, can be excited above the Fermi surface leaving a vacancy behind in the conduction band. This process creates a Fermi sea excitation constituting of a CB electron and a CB hole. The CB hole remains in proximity with the CB electron through scattering with electrons occupying the states below the Fermi surface[8]. The overall result of this interaction between the photoexcited exciton and the cloud of CB electrons and CB holes is the generation of a quasiparticle called exciton polaron. In most of the existing reports, the role of the CB hole is neglected due to its relatively bigger spatial size, leading it to be considered less significant in explaining exciton polarons.

The zero-momentum exciton polaron state can be approximated as a coherent superposition of the bare exciton ($\chi_{\mathbf{0}}^{\dagger}$) and the polaronic dressing of the exciton by the Fermi sea electron ($e_{\boldsymbol{j}}^{\dagger}$) and hole ($e_{\boldsymbol{k}}$) from the conduction band[9]:

$$P_{\mathbf{0}}^{\dagger} = \alpha \chi_{\mathbf{0}}^{\dagger} + \sum_{\boldsymbol{jk}} \beta_{\boldsymbol{jk}} \chi_{\boldsymbol{k}-\boldsymbol{j}}^{\dagger} e_{\boldsymbol{j}}^{\dagger} e_{\boldsymbol{k}}$$

with $|\boldsymbol{j}| > k_F$, $|\boldsymbol{k}| < k_F$, and $k_F = 2\sqrt{\pi N}$ is the Fermi wave vector. This interaction results in two energetically split eigenstates: (1) a lower energy attractive polaron, in which the exciton and Fermi sea excitations form a bound state, and (2) the higher energy repulsive polaron, which is a renormalized exciton state due to a net repulsive interaction between the exciton and the Fermi sea. While there has been a significant development in understanding the theory of exciton polaron physics, several aspects remain unclear, such as, (a) experimental signatures that clearly distinguish trion from exciton polaron, and (b) the role of the CB hole in the formation of exciton polaron.

To address the above questions, in this work, we study the interaction of confined excitons with a remote Fermi sea that lies closer to the electron side of the exciton. This is achieved through a lateral $N^+$-I-$P^-$ device geometry using a stack of monolayer $MoS_2$ placed on twisted-hBN (see **Methods** for device fabrication). We show that this stack confines the intra-layer excitons of $MoS_2$ in a sharp 1D potential well, as evidenced from quantized exciton emission. The spatially confined exciton possesses a non-zero permanent in-plane dipole moment, which interacts with a remote Fermi sea present on the electron side of the polarized exciton. Conventionally, a net repulsive interaction between the free electrons and the exciton refutes the possibility of formation of a trion bound state. However, similar to quantized excitons, we observe emission from quantized bound charged excitons. The level splitting between the trions and excitons is similar. We explain this observation in the framework of exciton polaron, with the necessary inclusion of a CB hole through an induced polarization of the Fermi sea by the exciton.

We stack two different flakes of hBN in the AA stacking order, ensuring a $\sim 0^{\circ}$ twist angle between the two flakes. The stack undergoes atomic reconstruction from AA stacking into relatively lower energy configurations of AB and BA stackings, that forms triangular domains spanning the entire sample area[10–12]. The atomic rearrangement puts the B (N) atom in the top layer above the N (B) atom in the AB (BA) domain. The empty and filled $2p_z$ orbitals of the vertically aligned B and N atoms hybridize. The resultant hybridized $2p_z$ orbital of the N atom with an off-centred electron distribution has a nonzero dipole moment[13] (Fig. 1a). Hence, an electric polarization in the out of plane direction emerges at the interface of the two hBN layers, pointing in opposite directions in the complementary AB and BA domains. The successive domains of opposite polarization are separated by a sharp domain wall (Fig. 1b). For marginally twisted hBN layers, the domain wall can be as narrow as ~30 nm[14]. Due to opposite polarization in the adjacent domains, a strong in-plane electric field emerges as a result of the abrupt change in the charge polarity across the domain wall (Fig. 1c-d). A monolayer of $MoS_2$ is transferred onto the twisted hBN (t-hBN) stack (optical image in Fig. 1e), resulting in the imprinting of the ferroelectric domains present at the t-hBN interface onto the $MoS_2$ layer[15]. We keep the top hBN flake thin ($\sim$5 nm) in the t-hBN stack to strengthen the transfer of the electric polarization on the $MoS_2$ monolayer. The presence of electric polarization in the t-hBN and its transfer on $MoS_2$ is verified using the Kelvin probe force microscopy (KPFM) technique (see **Methods**). The KPFM image of the stack is shown in Fig. 1f (with zoomed-in views shown in Figs. 1g-h). In the KPFM image, the two colours represent the AB and BA triangular domains with opposite polarizations. The sharp change in potential at the domain boundary indicates the presence of a large in-plane electric field.

A cross-section view of the electric field distribution is shown in Fig. 1c. The portions of the $MoS_2$ flake that fall on top of the AB and BA domains experience a vertical, but opposite electric field. The effect of this is similar to electrostatic gating. Due to an opposite electrostatic gating in the adjacent domains, a strong in-plane field ($E_{||}$) emerges in $MoS_2$ across the domain wall. The $E_{||}$ peaks at the center of the domain wall and sharply decays away from it (Fig. 1d). Thus, overall, the $MoS_2$ monolayer experiences positive (negative) gating by the perpendicular electric field inside the AB (BA) domains, and an in-plane electric field in a narrow region around the domain wall.

The schematic in Fig. 2 shows the resulting band bending of the conduction and valence band edges of $MoS_2$ along the axis perpendicular to the domain wall (denoted by exciton centre-of-mass coordinate X, elaborated later). The band diagram is similar to an $N^+$-I-$P^-$ junction, where the AB and BA domains correspond to the heavily doped $N^+$ and weakly doped $P^-$ sides of the junction, respectively, and the domain wall makes for the narrow intrinsic region. Due to Fermi level pinning in $MoS_2$, the electron doped side can be heavily doped to $N^+$, while it is typically difficult to dope $MoS_2$ strongly p-type (see experimental drain current versus gate voltage plot in **Supporting Information 1**), hence the hole doped side remains very weakly doped ($P^-$). Due to this band bending, the electrons and holes are spatially separated and drift towards the opposite sides of the domain wall, thus the domain wall is depleted of free charge carriers.

However, the domain wall confines the neutral excitons in a narrow region due to in-plane Stark effect. As $|E_{||}|$ varies nonmonotonically in our sample with a sharp peak around the domain wall center, the Stark effect induced reduction in exciton energy is maximum around the same point. This forms a narrow 1D potential well that confines excitons, causing quantization of the exciton energy (Fig. 2, bottom panel). Note that, the in-plane polarizability of the exciton in monolayer is stronger than out-of-plane polarizability, hence one would readily expect a stronger Stark shift by the in-plane electric field compared with similar out-of-plane field[16]. Such 1D confinement of the neutral exciton on the domain wall, while keeping the unbound charge carriers depleted, creates an intriguing platform to study confined exciton and (remote) exciton polaron, as discussed below.

We perform photoluminescence (PL) measurements on the sample at 4 K using an excitation laser wavelength of 532 nm. A representative PL spectrum from a sample (with no hBN capping on top of $MoS_2$) is shown in the top panel of Fig. 3a. For comparison, a PL spectrum from a pristine hBN-capped $MoS_2$ is also shown in the bottom panel of Fig. 3c. The immediate observation from Fig. 3a is the splitting of the exciton peak into multiple levels in the twisted hBN sample. Strikingly, these quantized levels are nearly equally spaced in energy (Fig. 3b). A similar equal energy separation was observed at other spots on the sample (shown in **Supporting Information 2**). The quantized levels are separated by $\hbar\omega \approx 5$ meV, and the overall separation between the highest and the lowest energy exciton peak is $\sim$21 meV. Under harmonic well approximation, this suggests an overall confinement potential well depth of 21 meV + $\hbar\omega/2 \approx 23.5$ meV. Below we explain the splitting as a result of the exciton center-of-mass wavefunction confinement in the presence of the spatially varying $E_{||}$.

An exciton wavefunction is composed of two components: (1) a relative component denoting the electron-hole separation that determines the exciton binding energy, and (2) the center-of-mass (COM) component describing the overall motion of the exciton as one entity. Under no external perturbation, the COM and the relative part of the exciton Hamiltonian can be decoupled in a central potential and the total exciton wavefunction is thus obtained by separation of variables ($\psi_{ex} = \psi_{COM} \times \psi_{rel}$). In that case, the exciton COM wavefunction is similar to that of a free-particle ($\psi_{COM} \propto e^{i\boldsymbol{k}.\boldsymbol{r}}$).

The scenario differs in our sample as decoupling the relative and the COM component cannot be done in the presence of $E_{||}$. To understand this, we write the exciton Hamiltonian in the presence of $E_{||}$ as shown below (see derivation in **Supporting Information 3**):

$$H = -\frac{\hbar^2}{2M}\frac{\partial^2}{\partial X^2} - \frac{\hbar^2}{2\mu}\left[\frac{\partial^2}{\partial {x_r}^2} + \frac{\partial^2}{\partial {y_r}^2}\right] - \frac{q^2}{4\pi\varepsilon r} + q\int_{X-x_r}^{X+x_r} E_{||}(x)dx$$

Here $X(=\frac{m_e x_e + m_h x_h}{m_e + m_h})$ is the exciton COM coordinate. $m_{e(h)}, x_{e(h)}$ are the mass and the $x$ coordinate of the electron (hole), respectively. We define $X$ axis along the direction of $E_{||}$, perpendicular to the domain wall. The kinetic energy term in the orthogonal COM coordinate $Y$ is omitted in the above equation for convenience, as the exciton COM wavefunction is free along the domain wall. $x_r\ (x_e - x_h), y_r(y_e - y_h)$ represent the relative coordinates. $M, \mu$ are the COM and the reduced exciton mass, respectively.

The last term in eqn. 1 denotes the Stark effect component of the Hamiltonian. It varies as a function of $X$, with its magnitude being highest at the domain wall, and reduces sharply as we go away from the domain wall. This creates a 1D potential well and localizes the exciton COM wavefunction along $X$. Physically, this means that the probability of finding the exciton is restricted in the region very close to the domain wall, as opposed to its uniform distribution in the absence of $E_{||}$. This 1-D exciton confinement leads to the formation of multiple quantized levels[17] as observed in the PL spectrum in Fig. 3a,c.

To understand this further, we numerically solve the exciton Hamiltonian in eqn. 1. The COM wavefunction for the first two quantized levels of the lowest energy 1s exciton is plotted in Fig. 3d. The wavefunction distribution of the hole with respect to the electron in the relative $(x_r, y_r)$ space is also shown in Fig. 3e and 3f, at the highlighted $X$ position in Fig. 3d. $E_{||}$ induces a static dipole moment in the otherwise spherically symmetric 1s exciton. Note that, the induced static dipole moment is the highest for the maximally-confined lowest-energy state and reduces for the higher quantized levels.

Note that, the experimental observation of equally separated quantized levels is indicative of a harmonic (parabolic) confinement potential. While the predicted potential profile from the simulation is symmetric about the minimum, the functional form is not a parabolic one. This difference is indicative of smoothening of the 1D potential well, likely due to inhomogeneity present in the sample. An inhomogeneous dielectric environment on the top surface of $MoS_2$ modifies the electric field distribution, rendering the confining potential approaching Gaussian due to central limit theorem. A Gaussian profile (and for that matter, any smooth profile), to a first order, can be approximated by a parabolic function close to its minimum, where the function does not change too rapidly in space:

$$V(x) = -V_0 e^{-\alpha x^2} \approx -V_0(1 - \alpha x^2) \text{ for small } x.$$

An approximate parabolic confinement in the presence of inhomogeneities explains the equally separated peaks observed in this sample. We anticipate the emergence of such a parabolic confinement potential to be a generic effect of inhomogeneity, such as moiré and interfacial excitons.

To reduce the sample inhomogeneity and observe the splitting more clearly, we prepare an hBN-capped monolayer $MoS_2$ sample transferred onto a different part of the same t-hBN stack. The PL spectrum of the hBN-capped $MoS_2$ on t-hBN stack is shown in Fig. 3c. We observe only two clearly split exciton peaks, with a large energy separation of ~23 meV. A comparison between the spectral features with and without the top hBN layer suggests that the confining potential well for the excitons is wider in the sample without the top hBN, allowing energetically closely spaced excitonic states. This widening of the potential well, as discussed above, is due to the smoothening of the potential well due to sample inhomogeneity. On the other hand, with top hBN capping, reduced inhomogeneity allows us to achieve a narrower potential well, enhancing the spectral separation of the confined excitonic states. Note that, the overall energy splitting achieved in this work is large as a result of efficient electrostatic imprint of the potential due to the small thickness of the top hBN flake in the twisted hBN stack.

Compared to the 1s exciton, the Stark shift becomes stronger as the exciton principal quantum number ($n$) increases[18]. We observe a larger peak splitting for the 2s exciton as compared to the 1s exciton in our sample (**Supporting Information 4**). This is because of the nonzero first order Stark shift for excitons with principal quantum number $n > 1$. To explain this, we take the example case of $n = 2$ excitons. The presence of an electric field breaks the radial symmetry of the central potential. The atomic orbitals with different angular momenta $l$ for $n = 2$ $(2s, 2p_x, 2p_y)$, that are otherwise orthogonal for a central potential, hybridize in the presence of an odd parity electric field operator. The electronic distribution of the hybridized $2s$, $2p$ orbital is off-centred, which gives it a net dipole moment. This enables a non-zero linear Stark shift resulting in a stronger spatial confinement of the COM wavefunction for excitons with $n = 2$. A simplified derivation of the hybridized orbitals and its wavefunction distribution is shown in **Supporting Information 4.**

We now turn our attention to the lower energy peaks, around the spectral range where typical charged exciton peaks are observed in $MoS_2$. Interestingly, at several spots, the spectrum shows two prominent charged exciton peaks (T1 and T2) (see Fig. 4a). Further, we observe that the energy spacing between T1 and T2 exactly mirrors the splitting observed in the higher-energy excitonic states X1 and X2. The energy separation for each pair (that is, X1-T1 and X2-T2) is ~35 meV. To further confirm the observation, we use near-resonant 633 nm excitation, which shows the two charged exciton peaks prominently (Fig. 4b). The observation of multiple exciton - charged exciton pairs suggests that each excitonic level confined in the domain wall forms its own charged exciton complex in our sample. Note that, we also see a background emission in the charged exciton region, which we attribute to the attractive polaron emission from the $N^+$ AB domain (see **Supporting Information 5** for further discussion).

As noted earlier, the domain wall region, where the excitons are localized due to Stark effect, $E_{||}$ depletes the free charge carriers. Hence the formation of a three-particle bound charged exciton (trion) is difficult to explain. The domain wall confined excitons can, of course, remotely interact with the sea of electrons from the $N^+$ region (we ignore presence of sufficient hole density in the $P^-$ region, see **Supporting Information 1**). However, as noted earlier, $E_{||}$ polarizes the confined excitons, providing them a non-zero static dipole moment. The direction of this polarization forces the electron to be closer to the electron sea (see Fig. 4c). Thus, the net interaction between the confined excitons and the electrons in the Fermi sea is repulsive in this case, refuting the possibility of formation of a bound three particle trion state. Therefore, to explain this observation, it is necessary to invoke a remote exciton-polaron picture (see Fig. 4d). We argue below that an attractive interaction between the confined polarized exciton and the Fermi sea is mediated through a conduction band (CB) hole.

In our sample, the confined excitons in the domain wall interact remotely with the Fermi sea excitations in the adjacent domain. The schematic of the lower energy attractive exciton polaron is shown in Fig. 4e. The permanent in-plane dipole moment of the confined exciton (with its electron facing the Fermi sea) polarizes the Fermi sea excitations in the adjacent region. To form a bound state between the polarized exciton and the CB electron-hole pair, the CB holes (electrons) is displaced closer to (away from) the domain wall. Thus, the presence of the CB hole helps generate a net attractive interaction for such charge arrangement, as shown in Fig. 4e. This rearrangement gives rise to the confined attractive polaron and corresponds to the lower energy spectral feature. The higher energy peaks, referred to as the excitonic peaks so far in this manuscript, thus correspond to the higher energy confined repulsive polaron branch. The near-equal separation between the two repulsive polaron peaks and the two attractive polaron peaks suggests similar binding energies for the two confined attractive polaron states.

To conclude, we observe an excitonic specie in which a linear chain is formed which consists of bound polarized electron-hole pair and a polarized Fermi sea excitation. Our results show the necessity of the inclusion of conduction band hole in the formation of quantized exciton polaron. This is an outcome of the charge rearrangement facilitated by the specific sample geometry. We expect similar quantized exciton polarons will be observable in other platforms where electrons and holes are spatially separated.

## Methods

### Sample Fabrication and Characterization

The hBN flakes were exfoliated from bulk crystal and two appropriate flakes with long sharp edges were identified in an optical microscope. The flakes were transferred on to Au/Ti (40/20 nm) substrate via pickup and drop technique using PC coated PDMS stamp. Edges of the hBN flakes were aligned near 0°. Subsequently, a monolayer of $MoS_2$ was exfoliated and transferred using a PDMS stamp. The sample was annealed at 300 degree C under high vacuum for 3 hours. A conductive AFM probe (Spark70 Pt) was used in non-contact mode for the KPFM measurement. A drive frequency of 3 kHz and a drive voltage of 3 V were used. The tip to sample distance was kept at 10-20 nm.

Photoluminescence measurements are taken in a closed-cycle cryostat at 4 K, using 532 and 633 nm laser excitation using an objective with a numerical aperture of 0.5.

**Potential Simulation**

Electrostatic simulation is performed by solving the steady-state Maxwell's equations in a 3D geometry. The geometry consisted of a monolayer of $MoS_2$ placed on top of two hBN layers. The top hBN thickness was kept at 5 nm. At the hBN-hBN interface, alternating triangles of opposing charge densities ($\pm$ 30 mC/m$^2$) were defined. All potentials are plotted with respect to infinity.

**Acknowledgement**

This work was supported in part by National Quantum Mission, an initiative of the Department of Science and Technology (DST), Government of India, a grant from Indian Institute of Science under IoE, a grant from DRDO, and a grant from I-HUB QTF, IISER Pune. K.W. and T.T. acknowledge support from the JSPS KAKENHI (Grant Numbers 21H05233 and 23H02052) , the CREST (JPMJCR24A5), JST and World Premier International Research Center Initiative (WPI), MEXT, Japan.

**References**

1. Efimkin, D. K. & MacDonald, A. H. Many-body theory of trion absorption features in two-dimensional semiconductors. *Phys. Rev. B* **95**, 1–10 (2017).

2. Huang, D. *et al.* Quantum Dynamics of Attractive and Repulsive Polarons in a Doped MoSe2 Monolayer. *Phys. Rev. X* **13**, 11029 (2023).

3. Suris, R. A. Correlation Between Trion and Hole in Fermi Distribution in Process of Trion Photo-Excitation in Doped QWs. *Opt. Prop. 2D Syst. with Interact. Electrons* 111–124 (2003) doi:10.1007/978-94-010-0078-9_9.

4. Sidler, M. *et al.* Fermi polaron-polaritons in charge-tunable atomically thin semiconductors. *Nat. Phys.* **13**, 255–261 (2017).

5. Efimkin, D. K., Laird, E. K., Levinsen, J., Parish, M. M. & Macdonald, A. H. Electron-exciton interactions in the exciton-polaron problem. *Phys. Rev. B* **103**, 1–15 (2021).

6. Liu, E. *et al.* Exciton-polaron Rydberg states in monolayer MoSe2 and WSe2. *Nat. Commun.* **12**, 1–8 (2021).

7. Van Tuan, D., Shi, S. F., Xu, X., Crooker, S. A. & Dery, H. Six-Body and Eight-Body Exciton States in Monolayer WSe2. *Phys. Rev. Lett.* **129**, 76801 (2022).

8. Van Tuan, D. & Dery, H. Turning many-body problems to few-body ones in photoexcited

semiconductors using the stochastic variational method in momentum space, SVM-k. *arXiv preprint*, arXiv:2202.08378 (2022). (2022).

9. Imamoglu, A., Cotlet, O. & Schmidt, R. Comptes Rendus Physique. *Comptes Rendus Phys.* **22**, 89–96 (2021).

10. Kim, D. S. *et al.* Electrostatic moiré potential from twisted hexagonal boron nitride layers. *Nat. Mater.* **23**, 65–70 (2024).

11. Zhao, P., Xiao, C. & Yao, W. Universal superlattice potential for 2D materials from twisted interface inside h-BN substrate. *npj 2D Mater. Appl.* **5**, 1–7 (2021).

12. Gilbert, S. M. *et al.* Alternative stacking sequences in hexagonal boron nitride. *2D Mater.* **6**, (2019).

13. Yasuda, K., Wang, X., Watanabe, K., Taniguchi, T. & Jarillo-Herrero, P. Stacking-engineered ferroelectricity in bilayer boron nitride. *Science (80-. ).* **372**, 1458–1462 (2021).

14. Woods, C. R. *et al.* Charge-polarized interfacial superlattices in marginally twisted hexagonal boron nitride. *Nat. Commun.* **12**, 1–7 (2021).

15. Kim, D. S. *et al.* Moiré ferroelectricity modulates light emission from a semiconductor monolayer. *Sci. Adv.* **11**, 1–8 (2025).

16. Pedersen, T. G. Exciton Stark shift and electroabsorption in monolayer transition-metal dichalcogenides. *Phys. Rev. B* **94**, 125424 (2016).

17. Thureja, D. *et al.* Electrically tunable quantum confinement of neutral excitons. *Nature* **606**, 298–304 (2022).

18. Lian, Z. *et al.* Stark Effects of Rydberg Excitons in a Monolayer WSe2 P-N Junction. *Nano Lett.* **24**, 3935–3941 (2024).

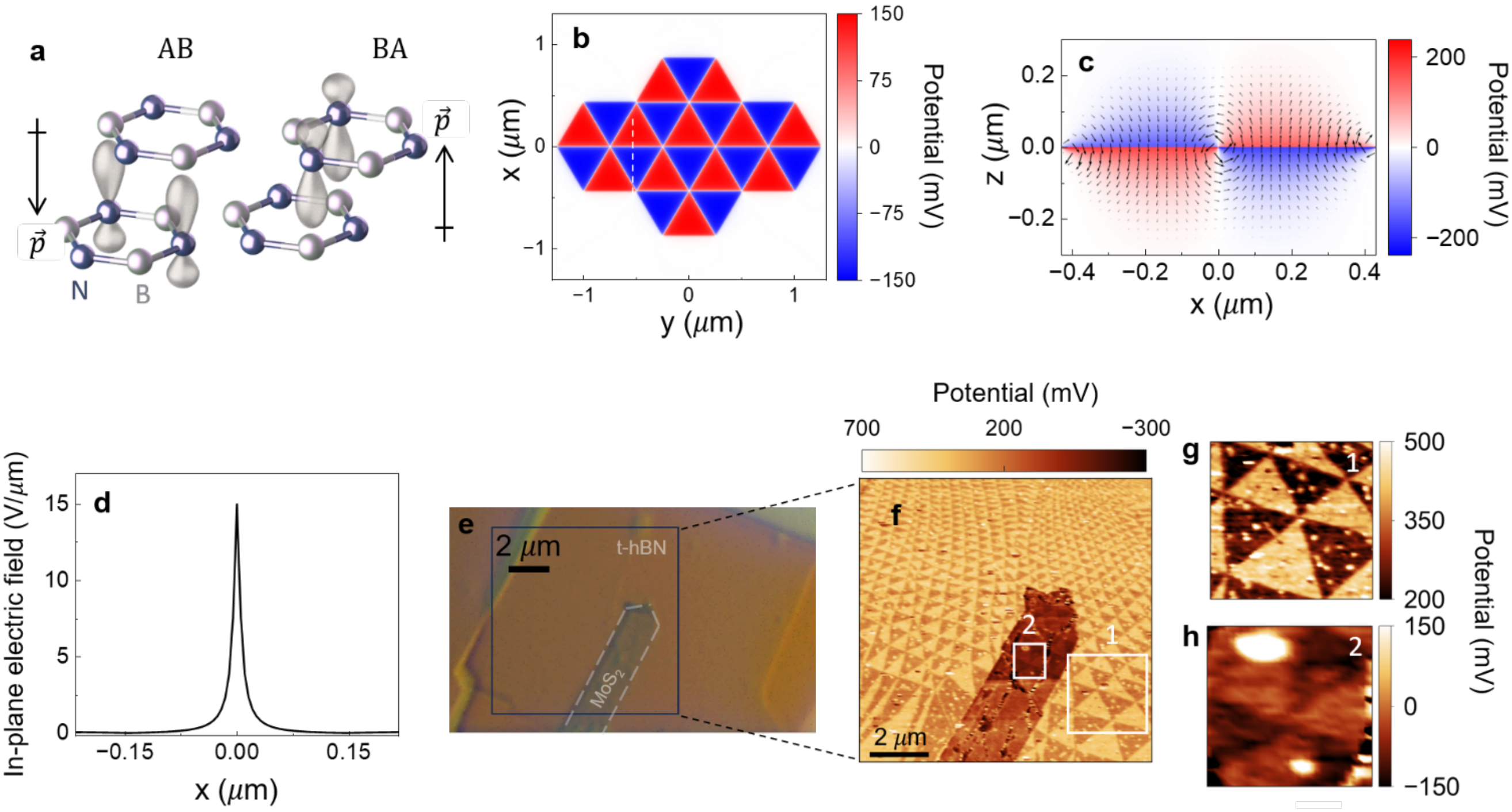

**Fig 1:** The interface of an AA-stacked hBN reconstructs atomically into lower energy AB and BA domains. (a) In these domains, the vertically aligned $2p_z$ orbitals of boron and nitrogen hybridize. The asymmetric distribution of the hybridized $2p_z$ orbital on nitrogen induces ferroelectricity along $z$, pointing in opposite directions in the two domains (represented by dipole moment $\vec{p}$). (b) Potential profile in the plane of $MoS_2$, as extracted from the electrostatic simulation of the $MoS_2$/t-hBN stack, showing imprinting of the ferroelectric domains from the t-hBN interface to $MoS_2$. (c) 2D cross-sectional view of the electric-field profile in the x-z plane, along the dashed white line in (b). Arrows indicate the magnitude and direction of the electric field. The plane of $MoS_2$ is $z = 0$ (the distance between $MoS_2$ and the t-hBN interface is taken to be ~5nm, similar to the top-hBN thickness in our experimental stack). Around the domain wall, the electric field is largely in-plane, pointing in the direction perpendicular to the domain wall. (d) Spatial variation of this in-plane electric field, along the white dashed line in (b). (e) An optical image of the fabricated $MoS_2$/t-hBN stack. (f-h) The emergence and imprinting of the ferroelectric domains is verified by the KPFM imaging, corresponding to the highlighted box in (e). Scale bar is 2 μm. Zoomed-in view of the boxes in (f) showing triangular domains in (g) t-hBN and (g) t-hBN/$MoS_2$ surface.

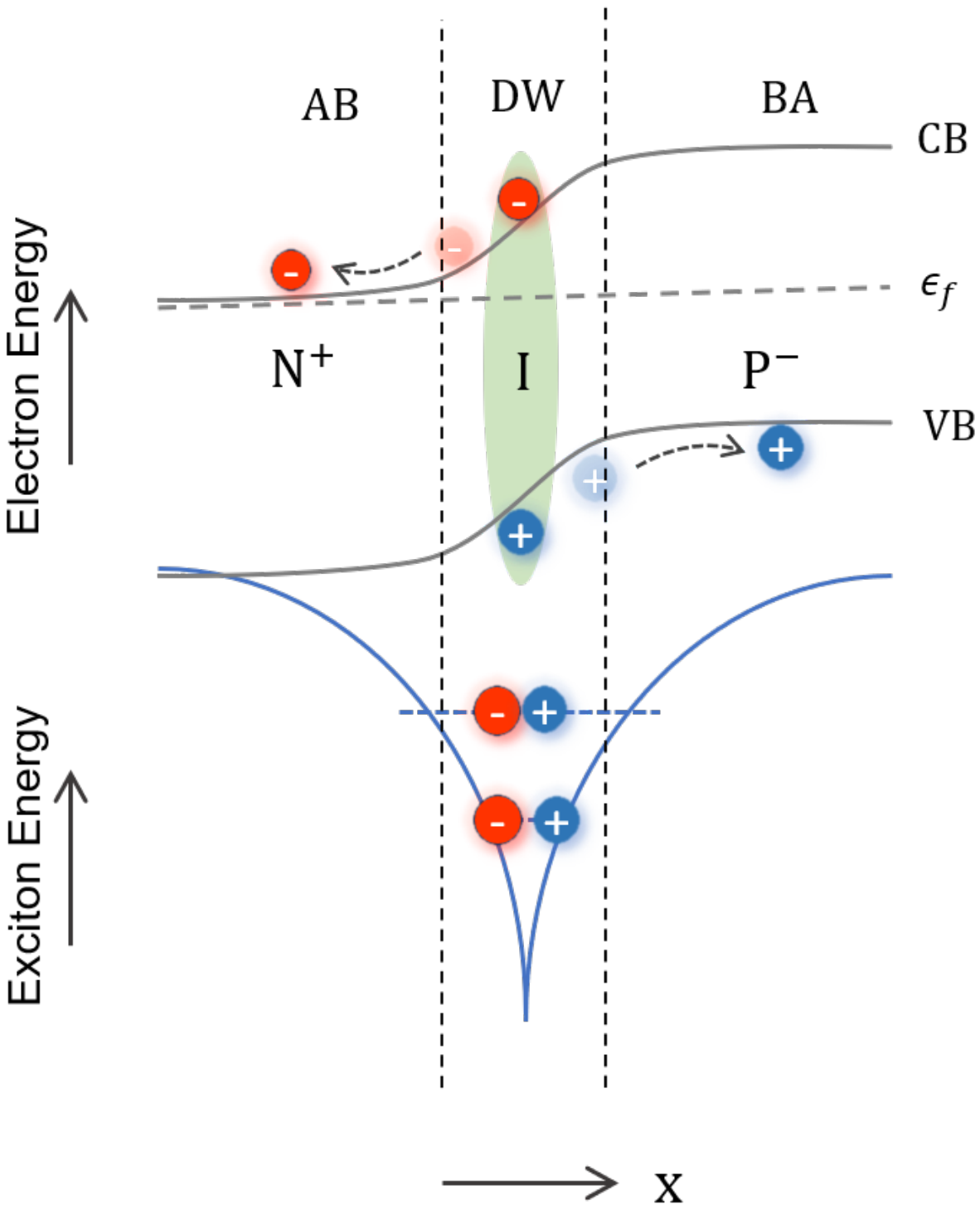

**Fig 2:** Top panel: The conduction and valence band bending in $MoS_2$ due to $E_{\parallel}$ which separates the electrons and holes into AB and BA domains. However, excitons are confined in the domain wall due to stark shift by $E_{\parallel}$. This results in the quantization of the exciton levels in the domain wall, and an induction of a net in-plane static dipole moment in the confined excitons due to $E_{\parallel}$. Bottom panel: Exciton confinement potential indicated by the blue curve.

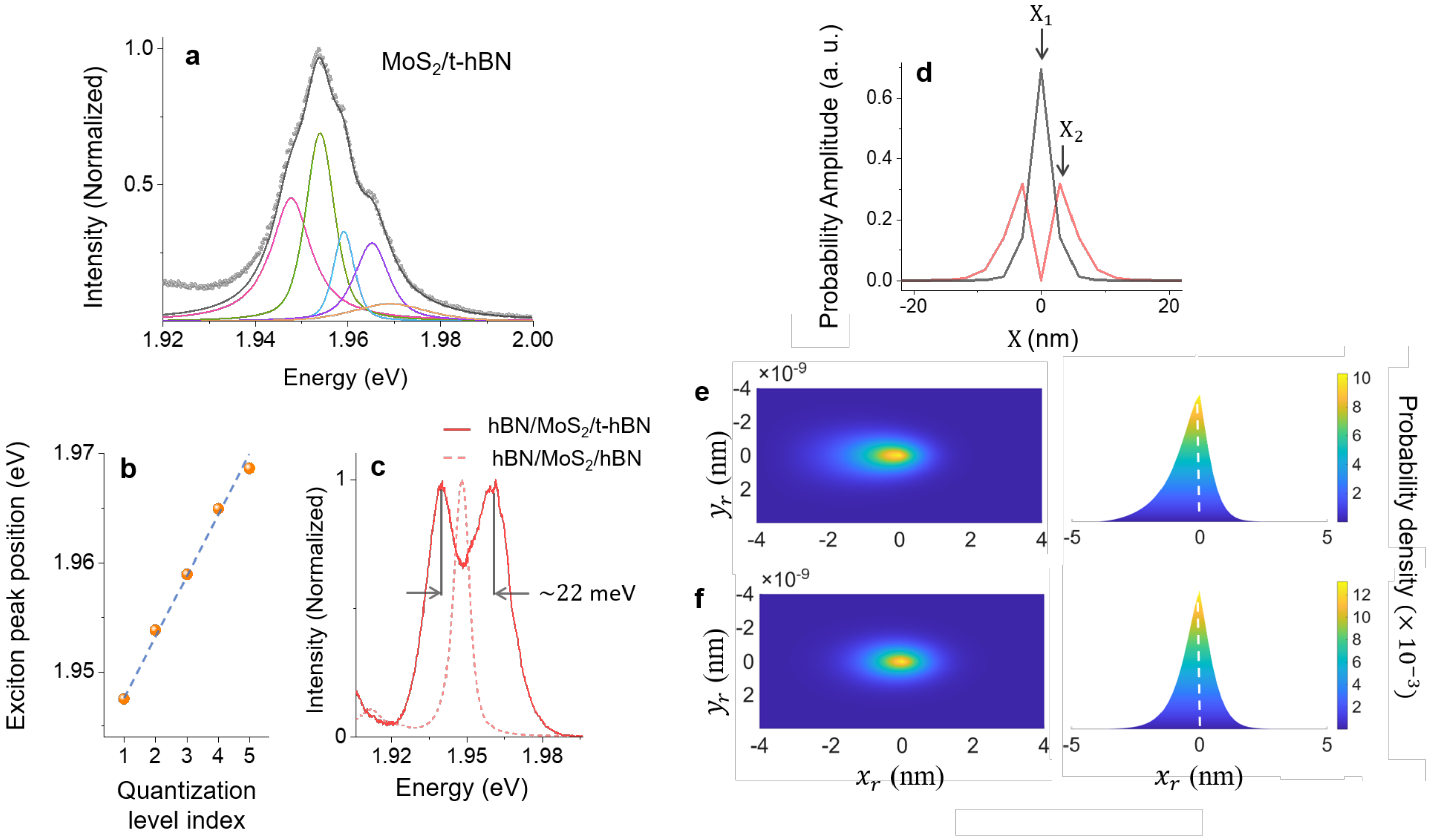

**Fig 3:** (a) PL spectrum of the $MoS_2$/t-hBN stack show multiple 1s exciton peaks due to Stark confinement by $E_{\parallel}$ in the domain wall. (b) A plot of exciton peak position versus the quantization index level, as extracted from (a). The linear fit indicates a nearly equal separation between the split energy levels. (c) PL spectrum showing ~22 meV Stark splitting in the hBN/$MoS_2$/t-hBN stack (in solid red). Reference data from hBN/$MoS_2$/hBN stack is also plotted for comparison (in dashed red). (d-f) The 1s exciton wavefunction as a function of (d) center-of-mass (COM), and (e-f) relative coordinates in the presence of $E_{\parallel}$. The COM wavefunction for the first (black) and the second (red) quantized levels in (d) is plotted along X (pointing perpendicular to the domain wall). The electron distribution relative to the hole (fixed at the origin) is shown for the (e) first and (f) second quantized levels at the COM coordinates $X_1$ and $X_2$, respectively. $X_1$ and $X_2$ correspond to the maxima of the exciton COM wavepacket distribution in (d). This asymmetric electron distribution about the hole center manifests as a net electric dipole moment, confirming exciton polarization. The polarization is strongest when the exciton is most confined and gets weaker as the confinement decreases for higher energy states.

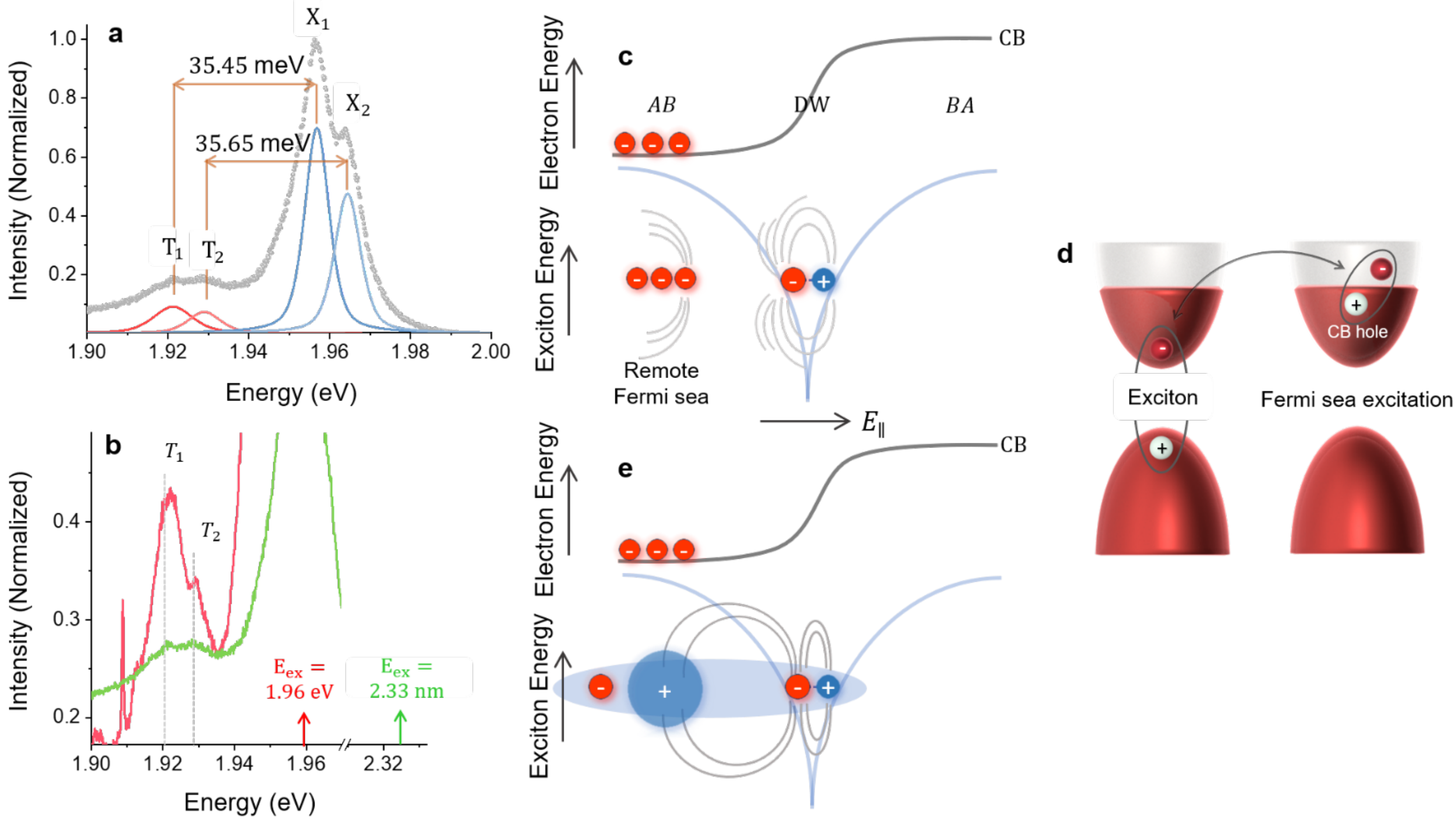

**Fig 4:** (a) PL spectrum, taken with 532 nm excitation, shows emission from charged ($T_1$ and $T_2$) and neutral excitons ($X_1$ and $X_2$). A similar energy separation is observed between $X_1$–$T_1$ and $X_2$–$T_2$, establishing $T_1$ and $T_2$ as the charged counterparts to the $X_1$ and $X_2$ exciton peaks. (b) The $T_1$, $T_2$ peaks appear more clearly under 633 nm excitation (in red) compared to 532 nm excitation (in green) due to near-resonant excitation. (c) Schematic of $E_{\parallel}$ forcing the electrons in the AB domain to be physically closer (farther) to the electron (hole) side of the polarized exciton. This implies a net repulsive interaction between the polarized exciton in domain wall with the electrons in the AB domain, thus forbidding the formation of a bound trion state. The picture of exciton polaron is invoked to explain the observation of $T_1$, $T_2$ peaks and its correlation with the quantized $X_1$, $X_2$ excitons lying in the domain wall, as observed in (a-b). (d) At high electron-doping, electrons in the Fermi sea, on interaction with the exciton, get excited above the Fermi surface and leaves behind CB holes. (e) Schematic illustrating that a charged exciton bound state in our stack can form through an attractive interaction between the polarized exciton and the polarized Fermi excitation. The sequential arrangement of the electron (CB) – hole (CB) - electron (exciton) – hole (exciton) forms the quasiparticle configuration of the attractive exciton polaron.

# Supporting Information:

# Confined exciton polaron in $MoS_2$ on twisted-hBN

Garima Gupta[1,2], Mayank Chhaperwal[1], Pankaj Kumar[1], Pushkar Dasika[1], Kenji Watanabe[3], Takashi Taniguchi[4], Archana Raja[2], and Kausik Majumdar[1,*]

[1]*Department of Electrical Communication Engineering, Indian Institute of Science, Bangalore 560012, India*

[2]*Molecular Foundry, Lawrence Berkeley National Laboratory, Berkeley, CA 94720, USA*

[3] *Research Center for Functional Materials, National Institute for Materials Science, 1-1 Namiki, Tsukuba 305-044, Japan*

[4] *International Center for Materials Nanoarchitectonics, National Institute for Materials Science, 1-1 Namiki, Tsukuba 305-044, Japan*

[*]*Corresponding author, E-mail: kausikm@iisc.ac.in*

### SI 1. Drain current versus gate voltage of $MoS_2$ transistor

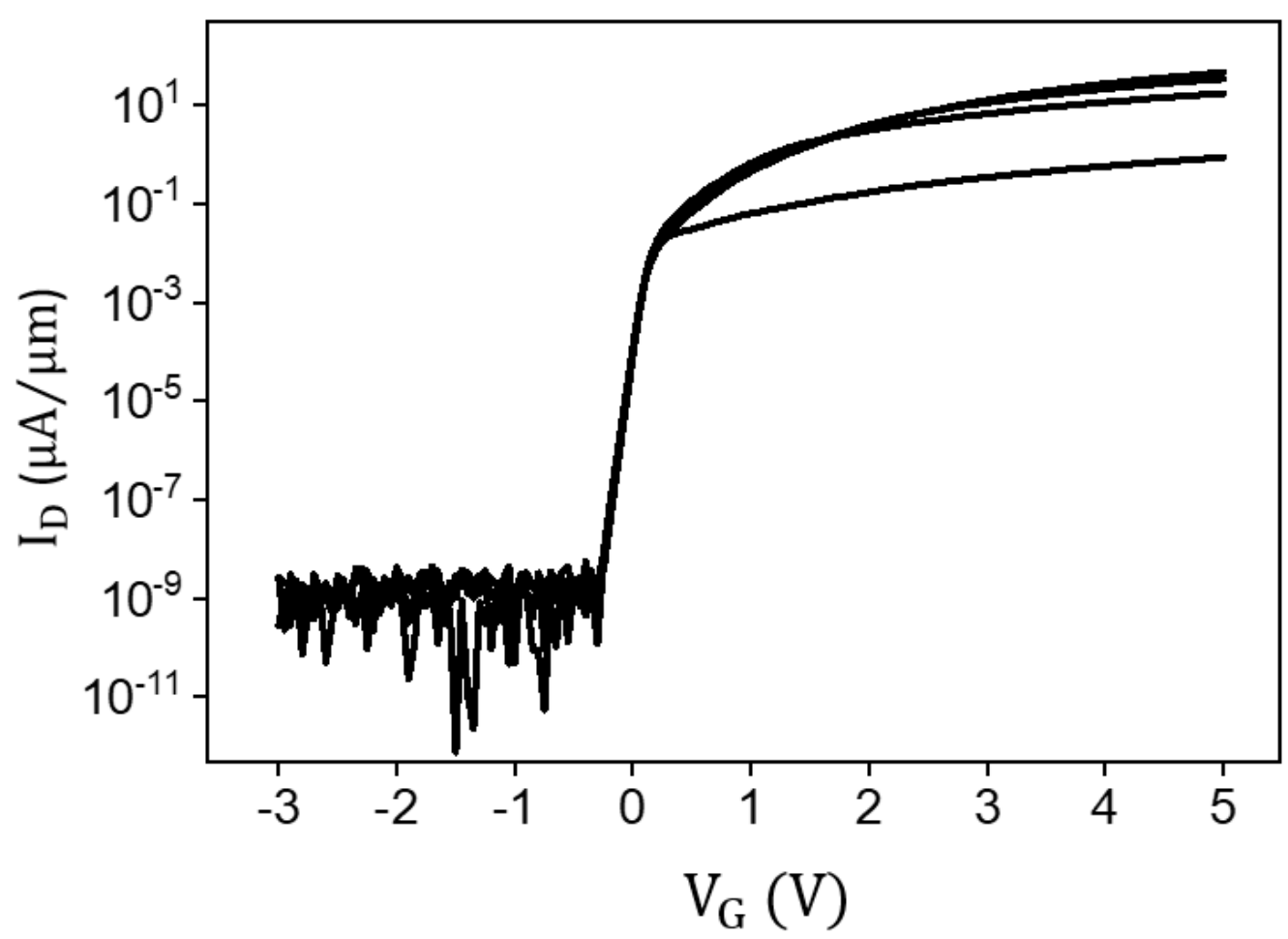

Fig. S1. Drain current versus gate voltage plot of monolayer $MoS_2$ taken on a different device. The individual curves correspond to different drain voltage values (0.05, 1.05, 2.05, 3.05 V). The low current in the device at negative gate bias, even up to -3V, suggests ineffective hole doping in $MoS_2$. This happens due to the pinning of the Fermi level in $MoS_2$ close to the conduction band.

**SI 2. Additional photoluminescence spectra from different spots of the uncapped sample**

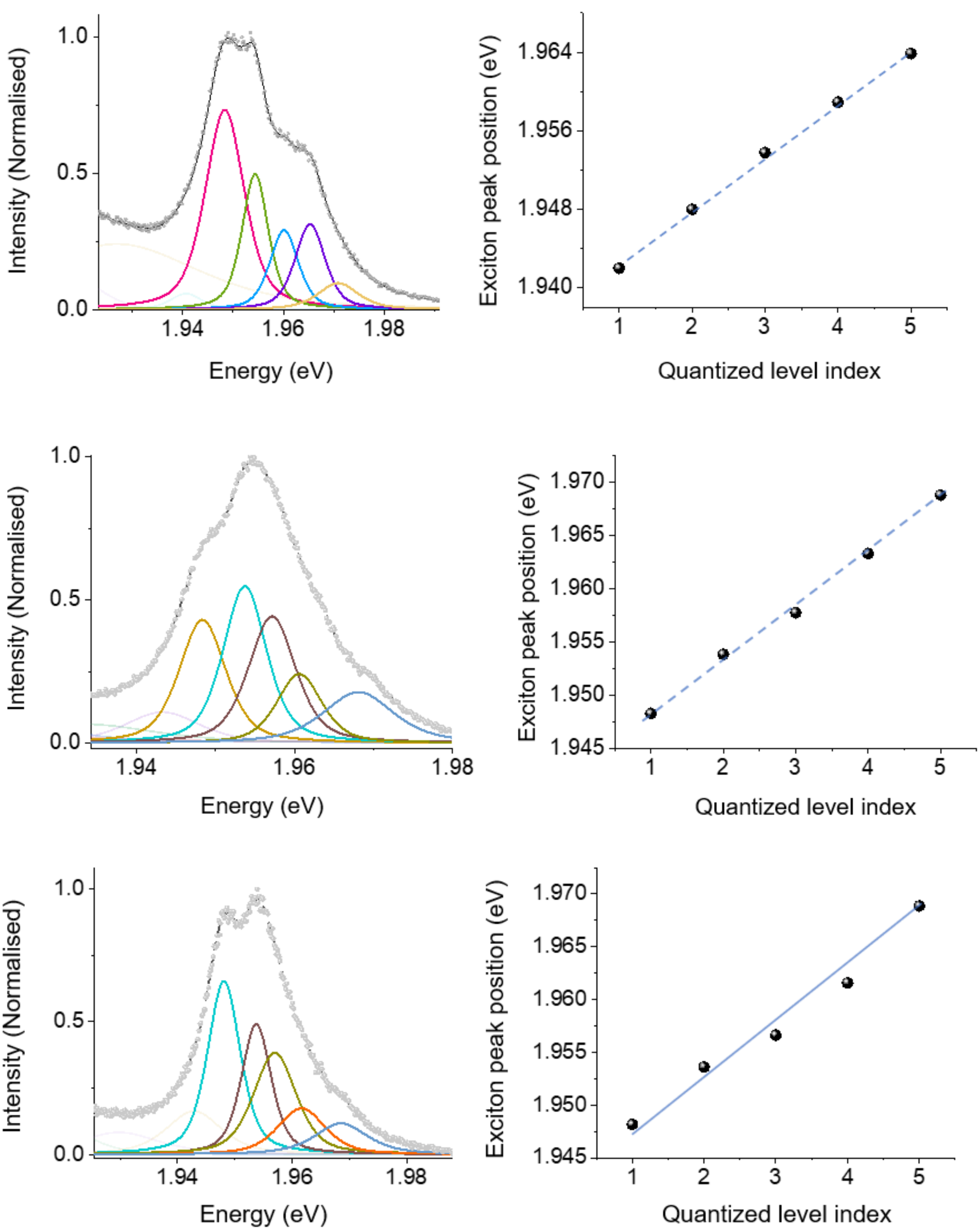

Fig. S2. PL spectrum of $MoS_2$ on t-hBN (without the top hBN) at different points on the sample (left column). The split exciton peaks obtained by spectral fitting are plotted in solid lines. The exciton peak position as a function of the quantized level index for each spectrum is also plotted (right column). The fitting of the peak energy values to a straight line indicates equally separated peaks.

## SI 3. Exciton Hamiltonian in a non-uniform electric field

The exciton Hamiltonian in a non-uniform electric field $[E(x)]$ is derived below. $x_{e(h)}, y_{e(h)}$ are the spatial coordinates of the electron (hole). $r$ is the relative electron-hole separation given by $r = [(x_e - x_h)^2 + (y_e - y_h)^2]^{1/2}$.

$$H = -\frac{\hbar^2}{2m_e}\left[\frac{\partial^2}{\partial {x_e}^2} + \frac{\partial^2}{\partial {y_e}^2}\right] - \frac{\hbar^2}{2m_h}\left[\frac{\partial^2}{\partial {x_h}^2} + \frac{\partial^2}{\partial {y_h}^2}\right] - \frac{q^2}{4\pi\varepsilon r} + q\left[-\int_{-\infty}^{r_h} E(x)dx + \int_{-\infty}^{r_e} E(x)dx\right]$$

The last term is the potential energy of electron and the hole in the presence $E(x)$.

$$H = -\frac{\hbar^2}{2m_e}\left[\frac{\partial^2}{\partial {x_e}^2} + \frac{\partial^2}{\partial {y_e}^2}\right] - \frac{\hbar^2}{2m_h}\left[\frac{\partial^2}{\partial {x_h}^2} + \frac{\partial^2}{\partial {y_h}^2}\right] - \frac{q^2}{4\pi\varepsilon r} + q\int_{r_h}^{r_e} E(x)dx$$

On changing the variables from electron/hole coordinate space to exciton centre of mass (COM) and relative coordinates, the above equation becomes:

$$H = -\frac{\hbar^2}{2M}\left[\frac{\partial^2}{\partial X^2} + \frac{\partial^2}{\partial Y^2}\right] - \frac{\hbar^2}{2\mu}\left[\frac{\partial^2}{\partial {x_r}^2} + \frac{\partial^2}{\partial {y_r}^2}\right] - \frac{q^2}{4\pi\varepsilon r} + q\int_{X-\frac{\mu}{m_h}x_r}^{X+\frac{\mu}{m_e}x_r} E(x)dx$$

Here $M(= m_e + m_h)$ and $\mu\ (= m_e m_h / m_e + m_h)$ represents the exciton COM and its reduced mass, respectively. $X\left(= \frac{m_e x_e + m_h x_h}{m_e + m_h}\right)$, $Y\left(= \frac{m_e y_e + m_h y_h}{m_e + m_h}\right)$ are the exciton COM coordinates, and $x_r(= x_e - x_h), y_r(= y_e - y_h)$ are its relative coordinates. We assume $m_e = m_h = m$ (that roughly approximately holds for monolayer TMDsMoS$_2$) to get:

$$H = -\frac{\hbar^2}{4m}\left[\frac{\partial^2}{\partial X^2} + \frac{\partial^2}{\partial Y^2}\right] - \frac{\hbar^2}{m}\left[\frac{\partial^2}{\partial {x_r}^2} + \frac{\partial^2}{\partial {y_r}^2}\right] - \frac{q^2}{4\pi\varepsilon r} + q\int_{X-x_r}^{X+x_r} E(x)dx$$

The solution of the above Hamiltonian will have a form $\psi(Y) \times \psi(X(x_r, y_r))$. The kinetic energy term corresponding to the free motion of the exciton COM in the Y direction can be solved independently and gives a free particle wavefunction of the form $\psi(Y) = e^{i\boldsymbol{k_y}.\boldsymbol{Y}}$. The Hamiltonian determining the constrained exciton wavefunction $[\psi(X(x_r, y_r))]$ is given below:

$$H = -\frac{\hbar^2}{4m}\frac{\partial^2}{\partial X^2} - \frac{\hbar^2}{m}\left[\frac{\partial^2}{\partial {x_r}^2} + \frac{\partial^2}{\partial {y_r}^2}\right] - \frac{q^2}{4\pi\varepsilon r} + q\int_{X-x_r}^{X+x_r} E(x)dx$$

We simulate the confined excitons in a domain wall by solving the Schrodinger equation $H\psi = E\psi$, using the above exciton Hamiltonian.

## SI 4. Stronger stark effect for excitons with principal quantum number $n = 2$

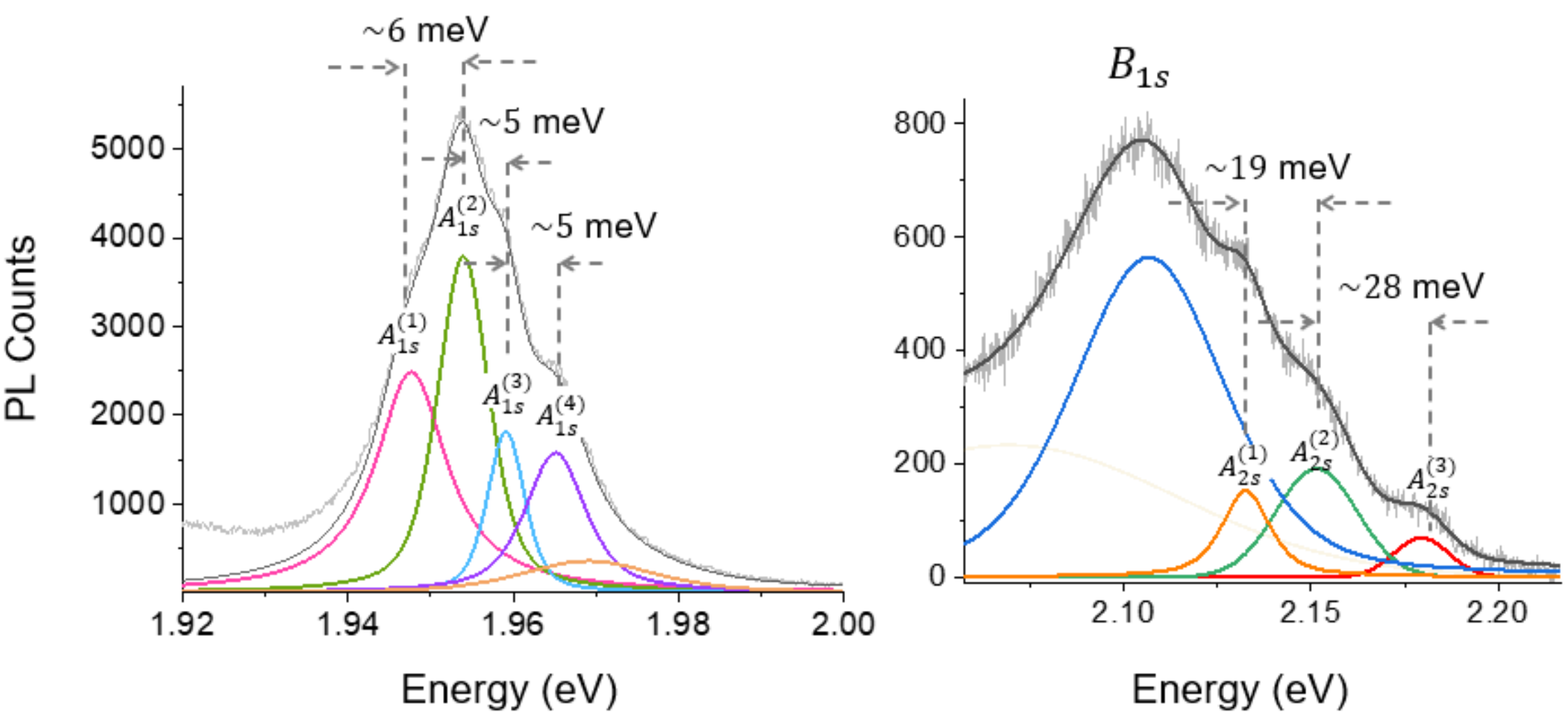

Fig. S3. This figure demonstrates the stronger Stark shift induced confinement of excitons in the domain wall of the $MoS_2$/t-hBN stack as a function of $n$. The raw and the fitted PL data is shown in grey and black solid lines, respectively. Left panel: The energy splitting for $n = 1$ excitons is $\sim 5 - 6$ meV. Right panelplot: The raw (in grey) and the fitted PL data (in black) is shown in the higher energy regime (corresponding to the PL spectrum shown in Fig. 3a in the main text)of $MoS_2$. The blue (fitted) peak corresponds to the $B_{1s}$ exciton emission. The higher three peaks are assigned to the quantized excitons in the $n = 2$ states showing higher energy splitting compared to the excitons in the $n = 1$ state. We rule out the assignment of these three peaks as coming from the quantization of the $B_{1s}$ exciton, as the exciton peak splitting in $B_{1s}$ can be expected to be in the same range as seen in the $A_{1s}$ exciton. The relatively larger separation of the $A_{2s}$ split peaks makes them visible in the emission spectrum.

To explain the observation of stronger Stark shift in $n = 2$ excitons, we take a simple case of an exciton in a uniform electric field. The perturbation due to $\vec{E} = \mathrm{E}\hat{x}$ is given by:

$$H' = -\vec{p}.\vec{E}$$

where $\vec{p}$ is the dipole moment vector given by $q\hat{r}$. $q$ and $\hat{r}$ denotes the electronic charge and the radial vector connecting the two charges, respectively. The perturbating Hamiltonian can be rewritten as:

$$H' = -q\hat{r}.\vec{E}$$

$$H' = -qrE\cos\varphi$$

where $\varphi$ is the in-plane azimuthal angle. The $n = 2$ excited state is three-fold degenerate $(2s, 2p_x, 2p_y)$ in a central potential. We represent these states by $|\psi_{2,l}\rangle$, where $l$ is the orbital angular momentum.

The first order correction in this threefold degenerate state due to $H'$ is obtained using the degenerate perturbation theory:.

$$H' = \begin{bmatrix} \langle\psi_{2,0}|\, H' |\psi_{2,0}\rangle & \langle\psi_{2,0}|\, H' |\psi_{2,1}\rangle & \langle\psi_{2,0}|\, H' |\psi_{2,-1}\rangle \\ \langle\psi_{2,1}|\, H' |\psi_{2,0}\rangle & \langle\psi_{2,1}|\, H' |\psi_{2,1}\rangle & \langle\psi_{2,1}|\, H' |\psi_{2,-1}\rangle \\ \langle\psi_{2,-1}|\, H' |\psi_{2,0}\rangle & \langle\psi_{2,-1}|\, H' |\psi_{2,1}\rangle & \langle\psi_{2,-1}|\, H' |\psi_{2,-1}\rangle \end{bmatrix}$$

Due to the odd parity of the perturbing Hamiltonian, the non-zero matrix elements in the above matrix exists only between states of opposite parities, i.e., with $\Delta l = \pm 1$. Hence, $H'$ is simplified as follows:

$$H' \propto \begin{bmatrix} 0 & 1 & 1 \\ 1 & 0 & 0 \\ 1 & 0 & 0 \end{bmatrix}$$

The new eigenstates of the system are $|\psi_\pm\rangle = \frac{1}{2}\left(\pm\sqrt{2}|2s\rangle + |2p_x\rangle + |2p_y\rangle\right)$, $|\psi_0\rangle = \frac{1}{\sqrt{2}}\left(|2p_x\rangle + |2p_y\rangle\right)$, the distribution of which is shown below.

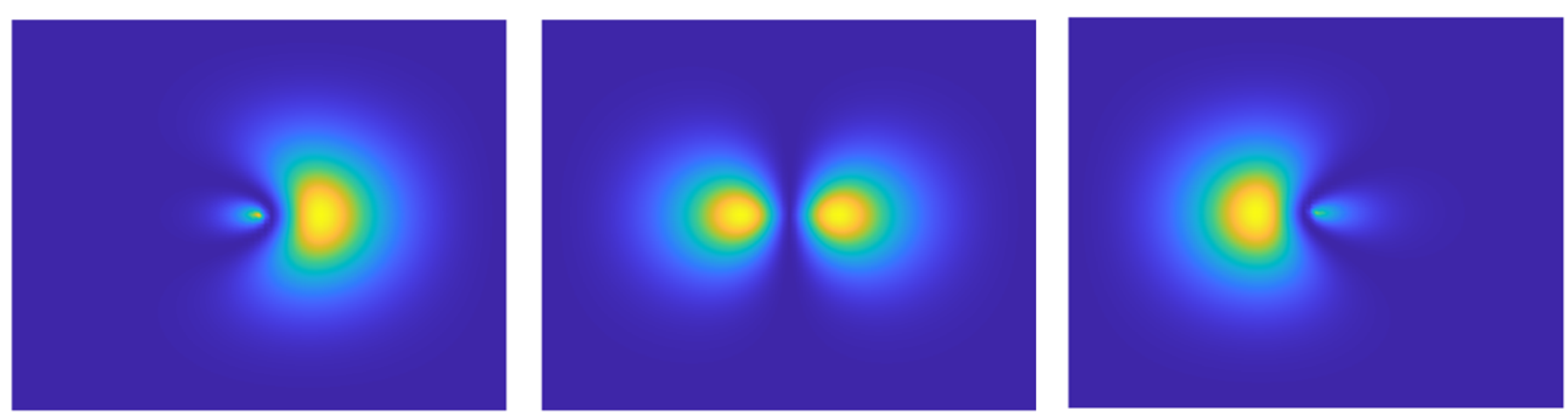

Fig. S4: The top-view of spatial distributions of the wavefunctions corresponding to the hybridized eigenstates.

As seen in the figure above, the charge distribution is asymmetric about the centre for the first and the third wavefunction distribution. Hence, the new mixed orbitals acquire a finite dipole moment. Hence, the presence of degeneracy distinguishes the linear (first order) stark effect in $n = 2$ from the $n = 1$ state, resulting in a stronger Stark effect induced peak shift[1].

### SI 5. Attractive polaron emission from the $N^+$ AB domain

At most of the spots on the sample, we observe a broad trion peak with no obvious splitting. We attribute this to the trion emission from the $N^+$ AB domain region in $MoS_2$, which dominates over the emission from the quantized exciton polarons (Fig. 4 in the main text). We assign the highest energy exciton peak observed to the continuum for the excitons in the domain wall. The $N^+$ domain trion together with this highest exciton state forms the attractive-repulsive polaron (AP-RP) pair for the exciton-Fermi sea system in the $N^+$ domain. We obtain the AP-RP peak separation and compare it as a function of local doping at different spots on our sample.

This trend mimics the blue shift of the RP peak with doping in gated samples, signifying exciton polaron formation[2] The plot below shows the dependence of the AP-RP peak separation (horizontal axis) on the corresponding RP peak position (vertical axis). Different points correspond to different spots on the sample. Due to spatial inhomogeneity, different points will correspond to different values of local doping. The data from a reference $MoS_2$/hBN sample is also highlighted in the same plot for comparison. The plot below shows that a higher trion binding energy (representing higher doping) leads to a larger blue shift of the RP peak. This trend mimics the blue shift of the RP peak with doping in gated samples, signifying exciton polaron formation[3]. This gives evidence of a relatively large electron doping in the $N^+$ AB domain of the $MoS_2$/t-hBN stack, overall pushing the towards a many-body exciton - Fermi sea interaction.

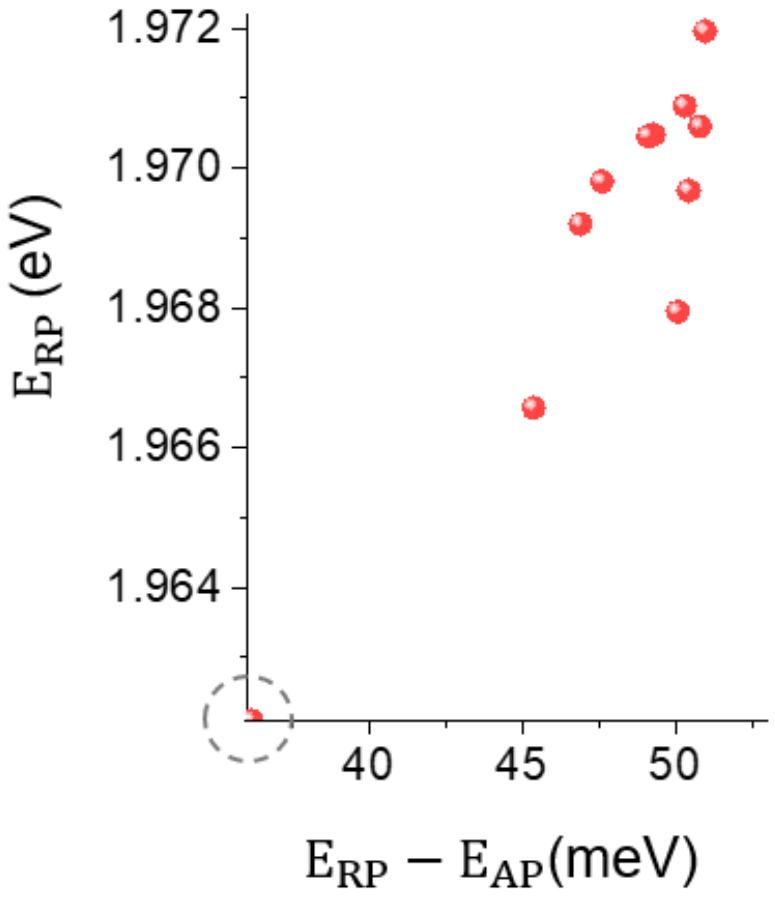

Fig. S5: Highest energy exciton plotted as a function its separation from the broad trion peak. Different points indicate different spots on the sample. The data point marked by the dashed circle corresponds to a a reference $MoS_2$/hBN sample.

## References

1. Lian, Z. *et al.* Stark Effects of Rydberg Excitons in a Monolayer WSe2 P-N Junction. *Nano Lett.* **24**, 3935–3941 (2024).

2. Efimkin, D. K. & MacDonald, A. H. Many-body theory of trion absorption features in two-dimensional semiconductors. *Phys. Rev. B* **95**, 1–10 (2017).

3. Huang, D. *et al.* Quantum Dynamics of Attractive and Repulsive Polarons in a Doped MoSe2 Monolayer. *Phys. Rev. X* **13**, 11029 (2023).